\documentclass[%
 aip,
 amsmath,amssymb,
 reprint,%
]{revtex4-1}

\usepackage{graphicx}% Include figure files
\usepackage{dcolumn}% Align table columns on decimal point
\usepackage{bm}% bold math
\usepackage[utf8]{inputenc}
\usepackage[T1]{fontenc}
\usepackage{mathptmx}
\usepackage{amsmath}
\usepackage{amsfonts}
\usepackage{amssymb}

\begin{document}

\preprint{AIP/123-QED}

\title[]{Correlation imaging through a scattering medium: experiment and comparison with simulations of the biphoton wave function}
% Force line breaks with \\

\author{Gnatiessoro Soro}
 \email{gnatiessoro.soro@femto-st.fr}

\author{Eric Lantz}%

\author{Alexis Mosset}%

\author{Fabrice Devaux}

\affiliation{Institut FEMTO-ST, D\'epartement d'Optique P. M. Duffieux, UMR 6174 CNRS \\ Universit\'e Bourgogne Franche-Comt\'e, 15b Avenue des Montboucons, 25030 Besan\c{c}on - France
}%

\date{\today}% It is always \today, today,

\begin{abstract}
We first extend our recent experiments of correlation imaging through scattering media to the case of a thick medium, composed of two phase scatterers placed respectively in the image and the Fourier planes of the crystal. The spatial correlations between twin photons are still detected but no more in the form of a speckle. Second, a numerical simulation of the biphoton wave function is developed and applied to our experimental situation, with a good agreement.
\end{abstract}

\maketitle

%\section{\label{Intro}Introduction}

Since its introduction in 1995 \cite{pittman_optical_1995}, correlation imaging based on entanglement has become a vivid field of quantum optics
\cite{moreau_imaging_2019}. In particular, it has been applied to image formation through inhomogeneous media, an important and challenging problem for which a wide variety of solutions has been offered \cite{kokhanovsky_light_2004}. In the last decades, imaging with entangled light through both a thin  \cite{peeters_observation_2010,defienne_adaptive_2018,gnatiessoro_imaging_2019} and a thick  \cite{lodahl_spatial_2005,skipetrov_quantum_2007,smolka_observation_2009,cande_transmission_2014} complex medium has been extensively considered. The spatial correlations hidden in the quantum  fluctuations of multiple scattered light have been investigated theoretically \cite{lodahl_spatial_2005,skipetrov_quantum_2007,beenakker_two-photon_2009,cande_transmission_2014} and experimentally \cite{smolka_observation_2009, gnatiessoro_imaging_2019}. In 2010, Peeters et al.\cite{peeters_observation_2010} reported experimental observation of two photon speckle patterns in the quantum correlations transmitted through either a surface or a volume scatterer using two independently scanning detectors. They measured coincidence count rates between punctual detectors.

The purpose of the present paper is twofold. First, we extend to multiple scattering light the results we have recently presented  \cite{gnatiessoro_imaging_2019} for a single phase diffuser, by imaging the spatial quantum correlations with two electron-multiplying charge coupled device (EMCCD) cameras. In the present experiment, the thick scattering medium consists in two identical phase diffusers with the second in the far-field of the first one, as in Ref. \cite{peeters_observation_2010}. Results obtained with two experimental arrangements are reported : one with the cameras placed in the image plane (i.e. near-field NF) of the thin crystal, and the second one with the cameras placed in the Fourier plane of the thin crystal (i.e. far-field FF). We also introduce  a numerical method of simulation of the biphoton wave function and compare its results to the experimental results presented here or obtained previously with a thin scattering medium \cite{gnatiessoro_imaging_2019}.

\begin{figure}[htbp]
\centering
\includegraphics[width=7cm]{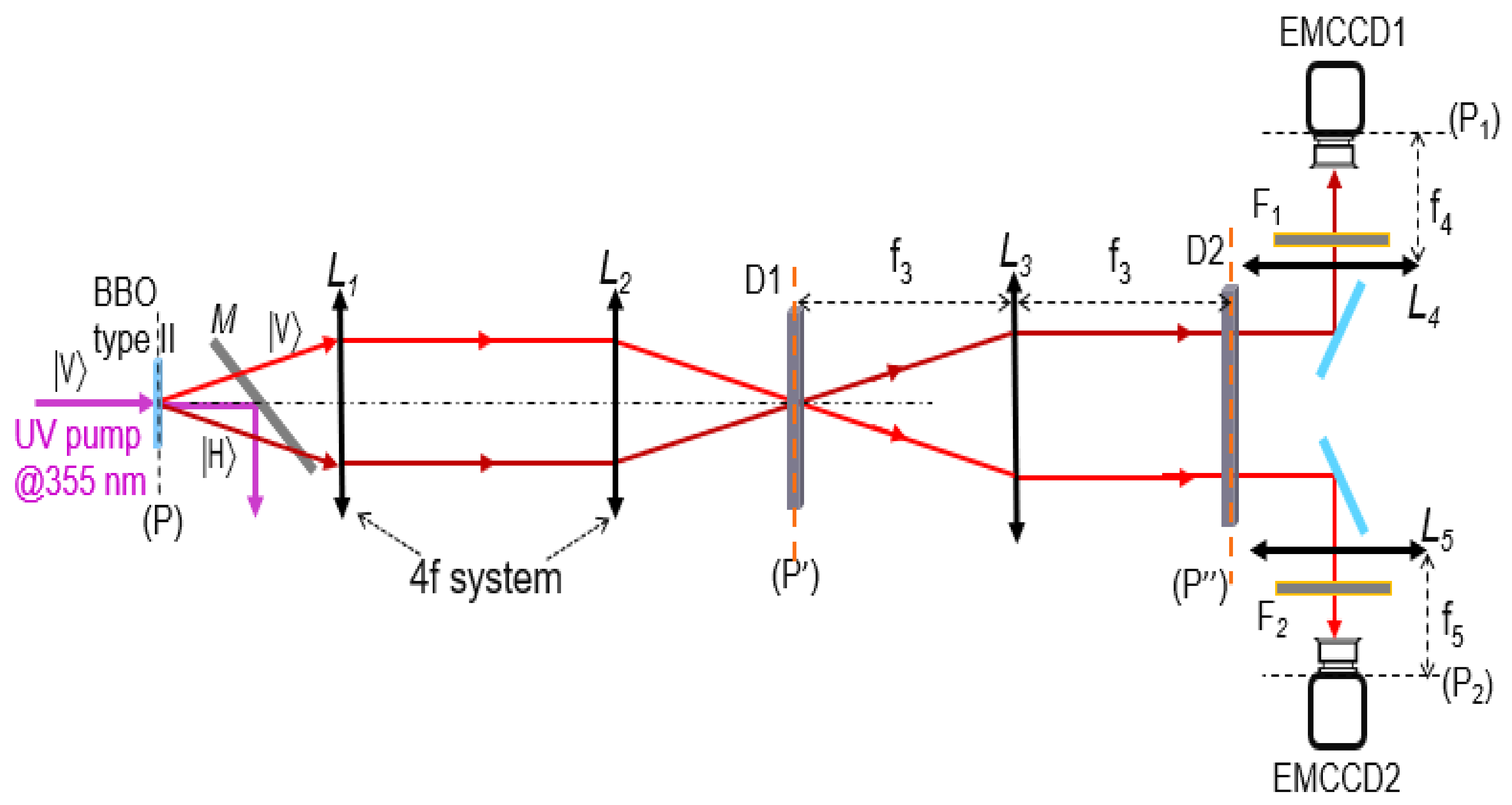}
	\caption{Experimental setup for NF correlations. First scatterer $D_1$ lies in the image plane $P'$ of the crystal and the second scatterer $D_2$ lies in the FF of $D_1$. Detection planes of the cameras are conjugated with $P'$}. 
\label{fig1}
\end{figure}

The first experimental setup is presented in Fig. \ref{fig1}. The collimated pulsed laser at 355 $nm$ (Q-switched Nd:YAG laser, 330 ps pulse duration, 27 $mW$ mean power, $1kHz$ repetition rate and 1.6 $mm$ FWHM beam diameter) illuminates a 0.8 $mm$ long $\beta$-barium borate (BBO) crystal. Entangled photon pairs are generated by noncolinear type-2 Spontaneous Parametric Down Conversion (SPDC). The $4f$ imaging system formed by lenses $L_1$ and $L_2$ ($f_1=f_2=125\,mm$) images the crystal (i.e. NF of the SPDC beams) onto the first thin scatterer $D_1$ with a magnification of 1. The cross-polarized entangled photon pairs transmitted by the scatterer $D_1$ are separated because of noncolinear phase matching and propagate through the second thin scatterer $D_2$ lying in the FF plane of the first one because of the lens $L_3$ ($f_3=75\,mm$). On both cameras, NF images of the SPDC beams are formed with a magnification of 2 by the afocal systems $L_3-L_4$ and $L_3-L_5$ ($f_4=f_5=150\,mm$) and with an exposure time of 100 $ms$ (i.e. 100 laser shots). The cameras are placed behind narrow-band interferential filters $F_1$ and $F_2$ ($@709\,nm$, $\Delta\lambda=4\,nm$), ensuring that the experiment operates in the quasimonochromatic regime. The thin scatterers are glass slides with one side attacked with fluorydic acid \cite{gnatiessoro_imaging_2019}.
\begin{figure}[htbp]
\centering
\includegraphics[width=8cm]{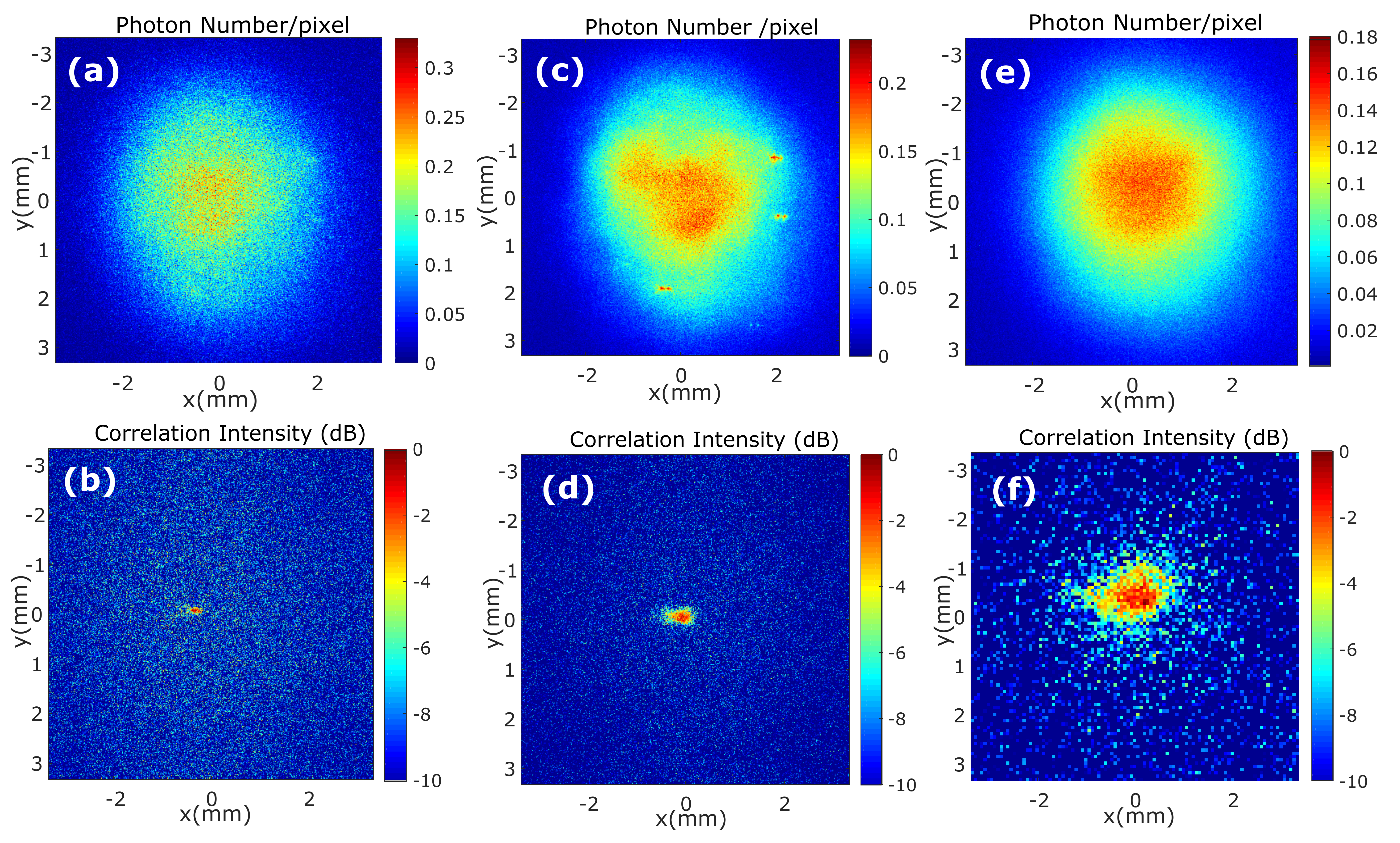}
\caption{Near-field experimental images. (a) and (b): without scatterer, (a) average photon number (signal or idler) (b) measured momentum correlation function in dB between 100 twin images. (c) and (d): with the thin scatterer $D_1$, (c) average photon number (d) measured momentum correlation function in dB over 40 000 twin images. (e)and (f): with the scatterers $D_1$ and $D_2$, (e) average photon number, (f) measured momentum correlation function in dB over 40000 twin images.}
\label{fig2}
\end{figure}
First, we discuss the experimental results when the two scatterers $D_1$ and $D_2$ are removed in the experimental setup. Fig. \ref{fig2}a shows a single NF image of the SPDC conditioned by the pump beam size. The spatial  position correlation function in Fig. \ref{fig2}b is calculated by summing the correlations between signal-idler images of the same pairs. While the spatial repartition of photons in the single NF image (Fig. \ref{fig2}a) is proportional to the pump beam profile, the position spatial correlation function shows a narrow correlation peak whose integral corresponds to a ratio of detection events in pairs \cite{devaux_quantum_2019} of 24\%. From figures \ref{fig2}a and \ref{fig2}b, and taking in account the magnification of 2 of the afocal systems, we measure the standard deviations of the SPDC beams $\sigma_{x}^{SPDC}=1.5\,mm$, $\sigma_{y}^{SPDC}=1.4\,mm$ and the standard deviations of the correlation peak $\sigma_{x}=8\,\mu m$, $\sigma_{y}=7\,\mu m$  along the $x$ and $y$ axes, respectively.

Second, we consider the case where the scatterer $D_1$ is inserted in the plane $P'$ without the scatterer $D_2$. Fig. \ref{fig2}c shows a single NF image of the SPDC transmitted through the scatterer $D_1$ and conditioned by the pump beam size. Because of the multimode character of the entangled light issued from one beam of the SPDC light, no one-photon speckle can be observed, as previously shown in Refs \cite{peeters_observation_2010,gnatiessoro_imaging_2019}. The position spatial correlation function obtained in Fig. \ref{fig2}d shows a narrow correlation peak with the ratio of detection events in pairs of 21\%. The ratio of the detection events obtained in Fig. \ref{fig2}b and Fig. \ref{fig2}d are close to the effective quantum efficiency 26\% of the entire detection system \cite{gnatiessoro_imaging_2019}. Indeed, the absence of a two-photon speckle pattern in the spatial correlation function in Fig. \ref{fig2}d is consistent with the fact that a pure phase object, even random, does not change the intensity in NF, with no effect on NF spatial correlations. The asymmetry observed in the spatial correlation functions (see figures \ref{fig2}b and \ref{fig2}d) is due to some residual astigmatism aberrations and defocusing of the imaging systems.

Finally, we discuss an experiment where the thick scatterer is used. The experimental set-up corresponds to Fig. \ref{fig1} where the scatterer $D_1$ lies in the plane $P'$ and the scatterer $D_2$ in the plane $(P^{\prime\prime})$. Fig. \ref{fig2}e shows a single NF image of the SPDC. The  spatial correlation function in Fig. \ref{fig2}f exhibits no speckle pattern, unlike with a single scatterer. On the other hand, the ratio of detection events in pairs is only slightly reduced to 16\%. We will see later in the article that the absence of structure in this correlation image can be explained by an incoherent superposition of speckle structures, different for each couple of signal-idler points. 
\begin{figure}[htbp]
\centering
\includegraphics[width=7cm]{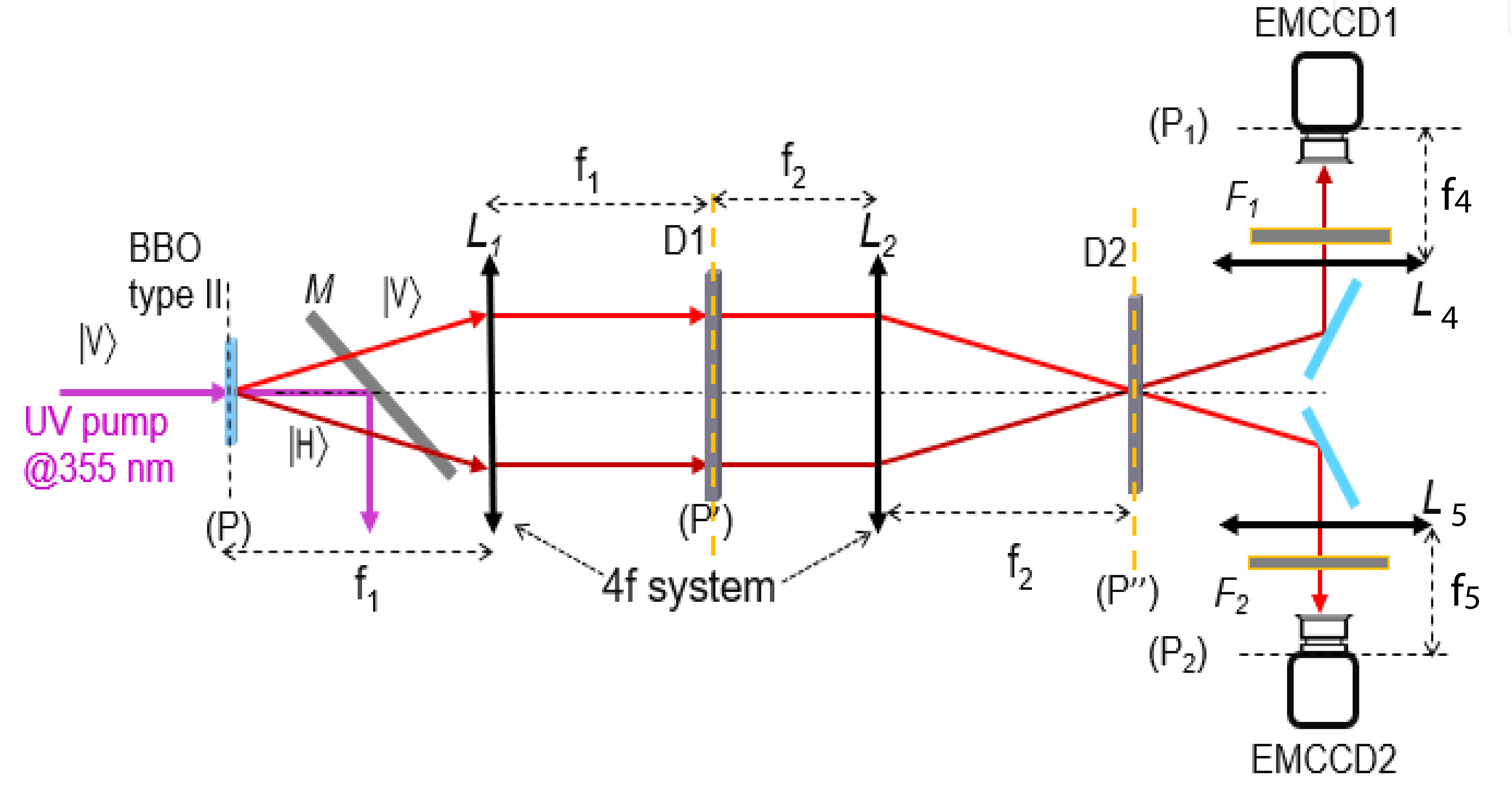}
\caption{Experimental setup for FF correlations. First scatterer $D_1$ lies in the Fourier plane $P'$ of the crystal and the second scatter $D_2$ lies in the FF of $D_1$. Detection planes of the cameras are conjugated with $P'$.}
\label{fig3}
\end{figure}
\begin{figure}[htbp]
\centering
\includegraphics[width=8cm]{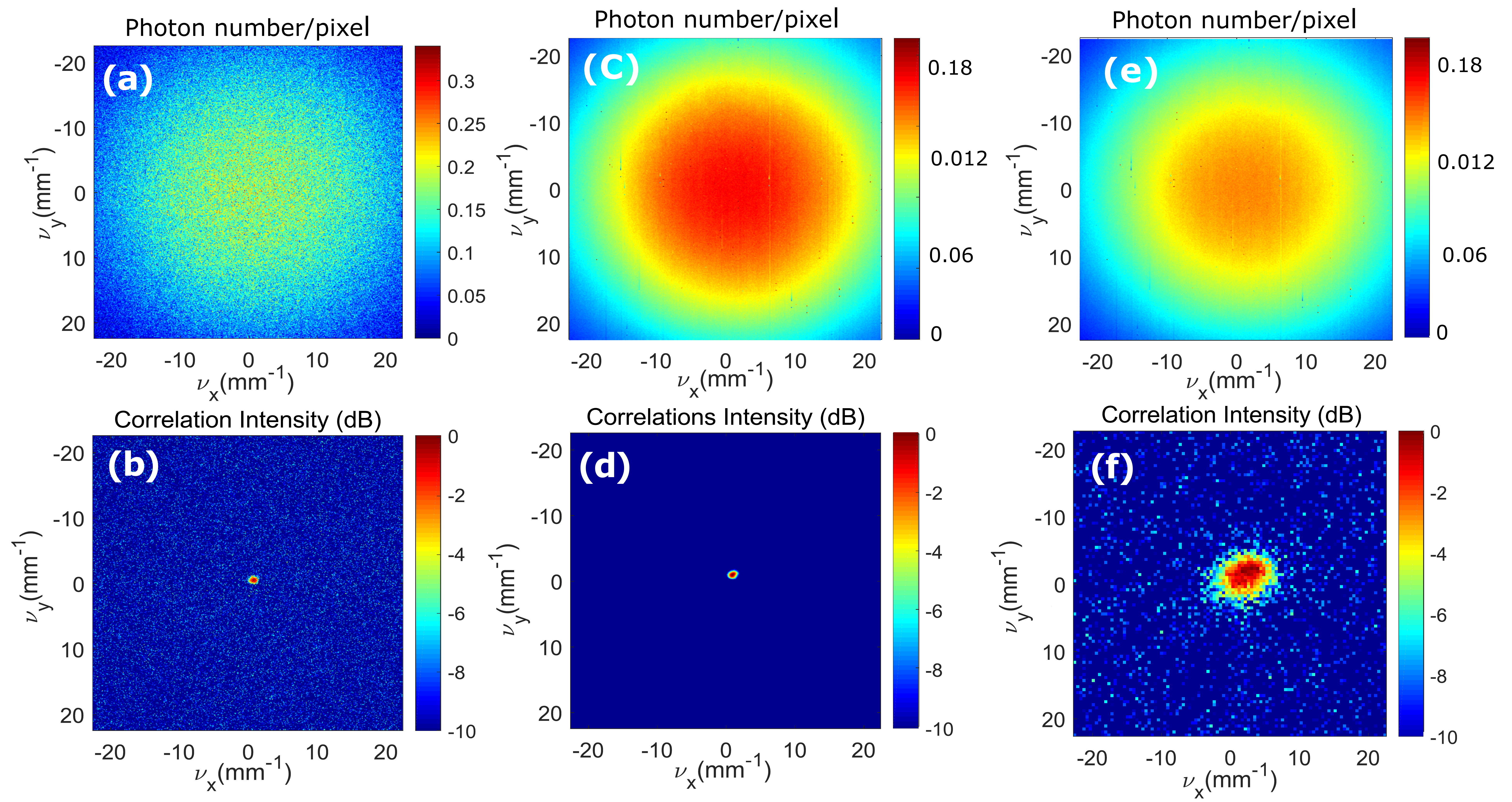}
\caption{ FF experimental images. (a) and (b): without scatterer, (a) average photon number (signal or idler) (b) measured momentum correlation function in dB between 100 twin images. (c) and (d): with the thin scatterer $D_1$, (c) average photon number (d) measured momentum correlation function in dB over 40 000 twin images.(e)and (f): with the sscatterer $D_1$ and $D_2$, (e) average photon number, (f) measured momentum correlation function in dB over 40 000 twin images.}
\label{fig4}
\end{figure}

While the previous experimental setup used to measure NF spatial correlations is similar to the configuration (c) used in ref. \cite{peeters_observation_2010}, the experimental setup used in this section and depicted in Fig. \ref{fig3} is different.
First, we remove the two thin scatterers $D_1$ and $D_2$ in the experimental setup and we measure the FF momentum  spatial correlations. Fig. \ref{fig4}a shows a typical single FF image of SPDC conditioned by phase matching. The spatial cross-correlation function obtained in Fig. \ref{fig4}b is calculated by summing the cross-correlations of the signal image  with the 180\ensuremath{^\circ} rotated idler image of the same pair of images. In this reference experiment, the momentum correlation in Fig. \ref{fig4}b shows a narrow peak. Similarly to the NF measurements, we calculate the ratio of the detection events in Fig. \ref{fig4}b and we obtain 21\%. From figures \ref{fig4}a and \ref{fig4}b, we measure the standard deviations (in spatial frequency unit) of the SPDC beams  $\sigma_{\nu_x}^{SPDC}= \sigma_{\nu_y}^{SPDC}=35\, mm^{-1}$ and of the correlation peak $\sigma_{\nu_x}=0.9\,mm^{-1}$, $\sigma_{\nu_y}=0.8\,mm^{-1}$ along the $x$ and $y$ axes, respectively. Considering the NF and FF spatial dimensions of the SPDC beams ($\sigma_{x}^{SPDC}\simeq\sigma_{x}^{pump}, \sigma_{y}^{SPDC}\simeq\sigma_{y}^{pump}$ and $\sigma_{\nu_x}^{SPDC}, \sigma_{\nu_y}^{SPDC}$) and their temporal properties (i.e. temporal and spectral standard deviations $\sigma_{t}^{SPDC}\simeq\sigma_{t}^{pump}=200\,ps$ and $\sigma_{\nu_t}^{SPDC}=\sigma_{\nu_t}^{F_1, F_2}=1.8\,THz$), we calculate the theoretical Schmidt numbers of the biphoton state along the $x$, $y$ and $t$ dimensions by:
\begin{equation} \label{eq1}
K_{x,y,t}=\frac{1}{2}\sigma_{x,y,t}^{pump}2\pi\sigma_{\nu_x,\nu_y,\nu_t}^{SPDC}
\end{equation}
Using Eq. \ref{eq1}, we obtain $K_x=165$, $K_y=154$ and $K_t=1.1\times 10^3$. Then, the spatio-temporal dimensionality of the entangled states defined as $K=K_{x}K_{y}K_{t}$ is approximately equal to $3\times10^7$.

Second, we consider an experiment where the scatterer $D_1$ is inserted in the plane $P'$ without the scatterer $D_2$. Figures \ref{fig4}c and \ref{fig4}d show a single FF image of SPDC and the spatial momentum correlation function, respectively. As for the NF spatial position correlations (see Fig. \ref{fig2}d ), Fig. \ref{fig4}d shows a narrow correlation peak, because we apply a pure phase object in the FF that does not modify the intensity used to measure momentum correlations. The degree of correlation (i.e the ratio of the detection events in pairs) is equal to 14\%. In this configuration, the degree of correlation with $D_1$ inserted is significantly reduced due to the propagation of the SPDC beams off the lens axis. In this case a significant part of the scattered light is not collected, particulary at $L_2$ lens.

Third, we consider the case of the thick scatterer where the scatterer $D_1$ lies in the plane $P'$ and the scatterer $D_2$ in the plane $P^{\prime\prime}$, as depicted in Fig. \ref{fig3}. In this case, Fig. \ref{fig4}e shows a single FF image of SPDC conditioned by phase matching. For the FF spatial momentum correlation function, Fig. \ref{fig4}f shows an incoherent superposition of speckle patterns with a FWHM of $\delta\nu_x\simeq\delta\nu_y=10\,mm^{-1}$. The ratio of the detection events in pairs calculated from Fig. \ref{fig4}f is equal to 12\%, close to that obtained with only the thin scatterer $D_1$ in Fig. \ref{fig4}d. The difference can be explained by supplementary losses induced by the second thin scatterer $D_2$. From the standard deviations of the NF and FF correlation peaks, we estimate the experimental Schmidt number or the degree of the Einstein-Podolsky-Rosen (EPR) paradox \cite{moreau_einstein-podolsky-rosen_2014} to $V=\sqrt{V_{x}V_y}=\sqrt{122\times264}\simeq 180$. This value is smaller than the spatial Schmidt number obtained with Eq. \ref{eq1} because of all imperfections of the imaging systems, especially for NF measurements where the correlation peak is enlarged because of astigmatism aberrations (due to off-axis propagation of the SPDC beams).  

We report now a purely numerical method of simulation of the biphoton wave-function and show results in good agreement with the experimental preceding sections. Let us first consider a thin crystal, illuminated by a pump beam with a transverse amplitude distribution $E_{p}(r)$. The  biphoton wave function at two points $r_s=(x_s,y_s)$ and $r_i=(x_i,y_i)$ of the detection plane for the signal and the idler, respectively,  can be written as \cite{saleh_duality_2000}:
\begin{equation} \label{thin}
|\psi(r_s,r_i)>=\int dr  E_p(r)h_s(r_s,r)h_i(r_i,r)
\end{equation}
where $h_s(r_s,r)$ and $h_i(r_i,r)$ are the impulse response functions of the linear, passive, non absorbing optical systems in which the fields propagate between the crystal and the signal/idler detection planes. Eq.\ref{thin} can be established by expressing the correspondence between the Schrödinger and the Heisenberg points of view for such a system \cite{bachor_guide_2004}. Note that the detection can occur in the Fourier plane, in which case the transverse coordinates $r_s$, $r_i$ are proportional to the spatial frequencies in the crystal plane.
  
Let us now consider a thick crystal. An analytical treatment becomes much more difficult and leads to double or triple integrals\cite{saleh_duality_2000, abouraddy_entangled-photon_2002}, if we attempt to let appear explicitly physical features like the phase matching function, an object in one or two beams and so on. For two signal and idler images of $N\times N$ pixels, the biphoton wave function includes $N^4$ values. If each value is calculated with a triple integral, the numerical computation evolves as the $10^{th}$ power of N, which is prohibitive. We propose here a purely numerical approach, that consists in writing the wave function for a thick crystal as the coherent sum of the wave functions corresponding to each slice of the crystal:
\begin{equation} \label{thick}
|\psi(r_s,r_i)>=\int dz\int dr  E_p(r,z)h_s(r_s,r,z)h_i(r_i,r,z)
\end{equation}
where $z$ stands for the coordinate in the crystal along the propagation direction. For a given crystal slice, the propagation in the further slices is taken into account in the Fourier domain by multiplying the slice impulse response by a quadratic phase term, in a way similar to the usual split-step propagation algorithm employed to solve, in classical nonlinear optics,  the nonlinear Schrödinger equation. The method is valid inasmuch as it is possible to neglect the chance that a pair of twin photons seeds the production of a further pair. This approximation is inherent to the  formalism of the biphoton wave function: it is assumed that all pairs are independent. This is the condition to obtain a whole description of  parametric fluorescence by signal-idler joint probabilities describing a unique pair of photons. This condition is fulfilled if the gain per spatio-temporal mode is much less than unity. This approximation, inherent to the very basis of the formalism, is the only one: all geometries can be envisioned, including non homogeneous (periodically poled) crystals \cite{t-mills_simulating_2020}, multiple crystals, and, here, a diffusing medium after the crystal.
  
With this method and for $N\geq32$, the computation time is proportional to $M\times N^6$, where $M$ is the number of crystal slices. For each slice, $N^2$ impulse responses are computed, by propagating an input numerical Dirac pulse (unity on one pixel, zero on the others). For $N\geq32$, the computation time of these impulse responses is negligible versus the time of computation, by a simple term by term multiplication, of the $N^4$ contributions to the wave function resulting, for one input Dirac pulse, from the two output fields.  With $M=40$ crystal slices, sufficient for the $0.8 mm$ BBO crystal of the experiment, the computation time on a professional PC is about 8 minutes for an image of $32\times32$ pixels, 8 hours for $64\times64$ pixels. The subsequent steps are rapid and straightforward. First, the non normalized signal-idler joint probability is computed as the squared  modulus of the non normalized wave function. Then, the quantities of interest are obtained by the appropriate summations. Two examples are as follow. The signal intensity for a pixel results from a sum of the $N^2$ joint probabilities between this signal pixel and all the idler pixels. The correlation for a given value of $r_s-r_i$ is obtained by summing all the joint probabilities corresponding to this difference and dividing by the total signal (or idler) intensity. With this definition, the correlation is normalized: we have verified that the sum of its values over all pixels is equal to one. The other quantities of interest are calculated in a similar way. 
\begin{figure}[htbp]
	\centering
	\includegraphics[width=6cm]{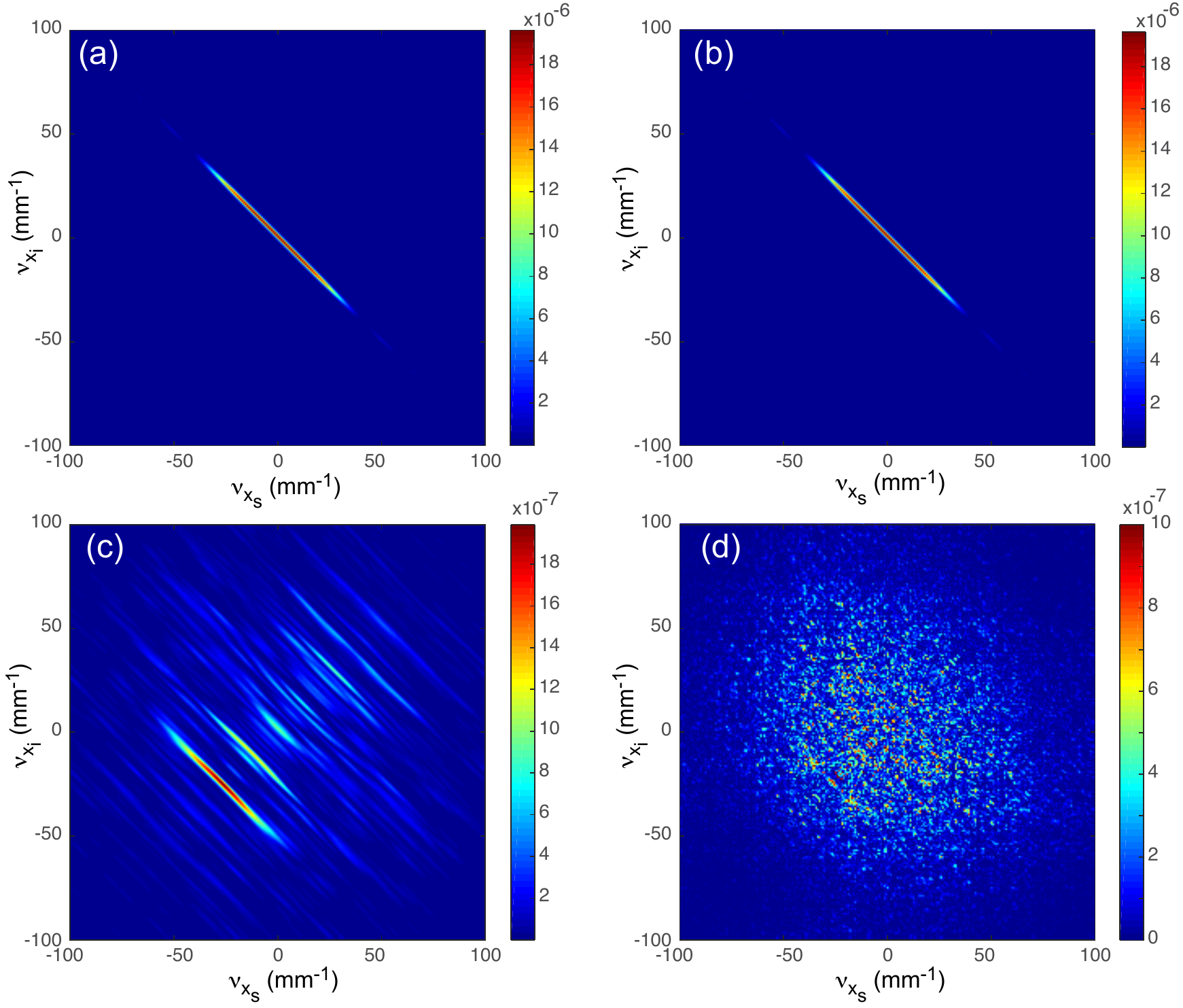}
	\caption{Far-field 1-D signal-idler correlations for a type 1 crystal (a) no diffusing medium (b) a scatterer in the FF  (c) a scatterer in the NF (d) scatterers both in the NF and in the FF.} \label{unD}
\end{figure}
\begin{figure}[htbp]
	\centering
	\includegraphics[width=8cm]{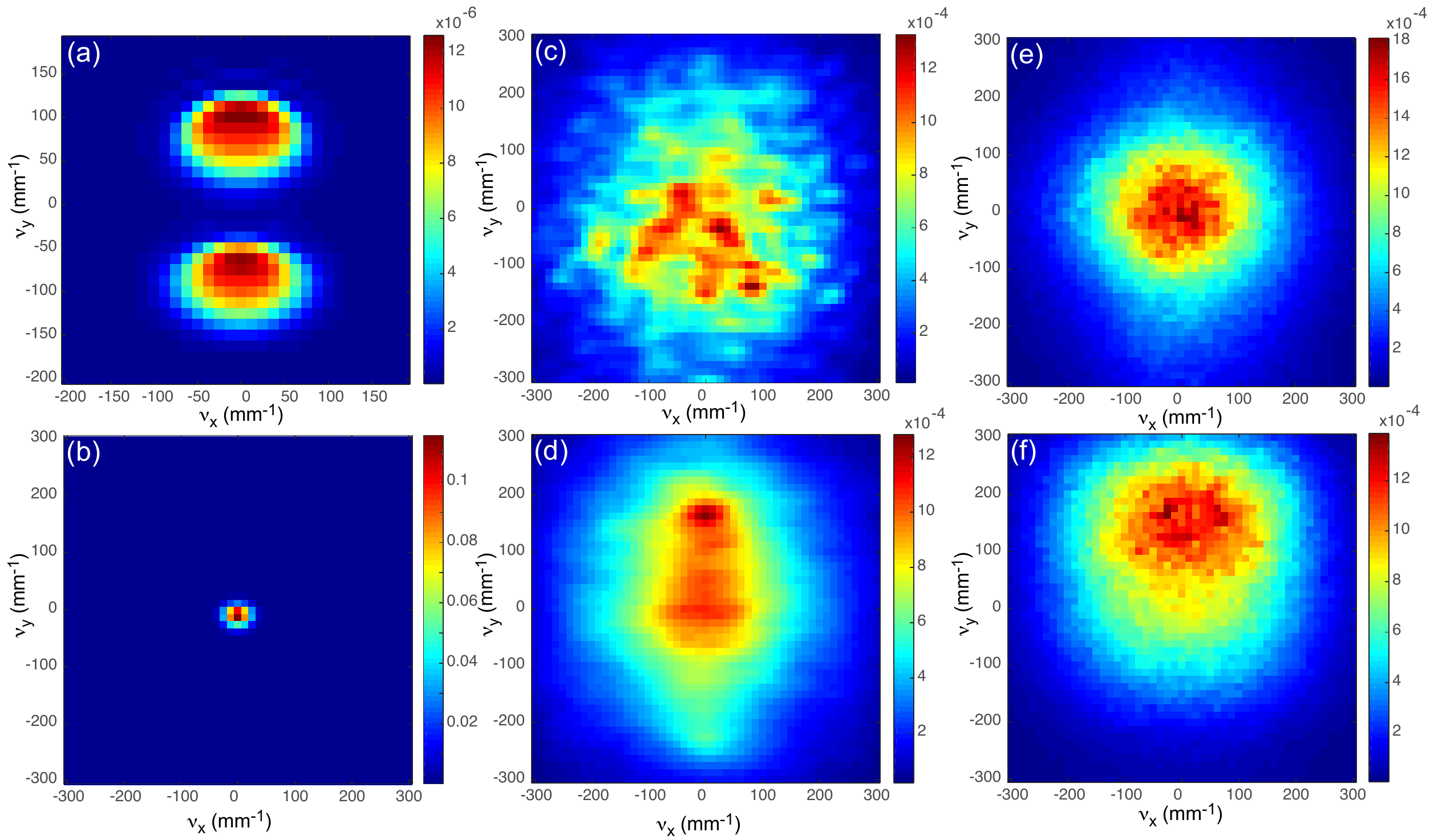}
	\caption{Far-field 2-D signal-idler intensities and correlations for a BBO crystal. (a) sum of the signal and idler intensities (b) $r_s+r_i$ correlation without scatterer (c) $r_s+r_i$ correlation with a scatterer  in the NF (d) $r_s-r_i$ correlation with a scatterer in the NF  (e) $r_s+r_i$ correlation with scatterers  in the NF and in the FF (f) $r_s-r_i$ correlation with scatterers in the NF and in the FF.}\label{twoD}
\end{figure}
\begin{figure}[htbp]
	\centering
	\includegraphics[width=4cm]{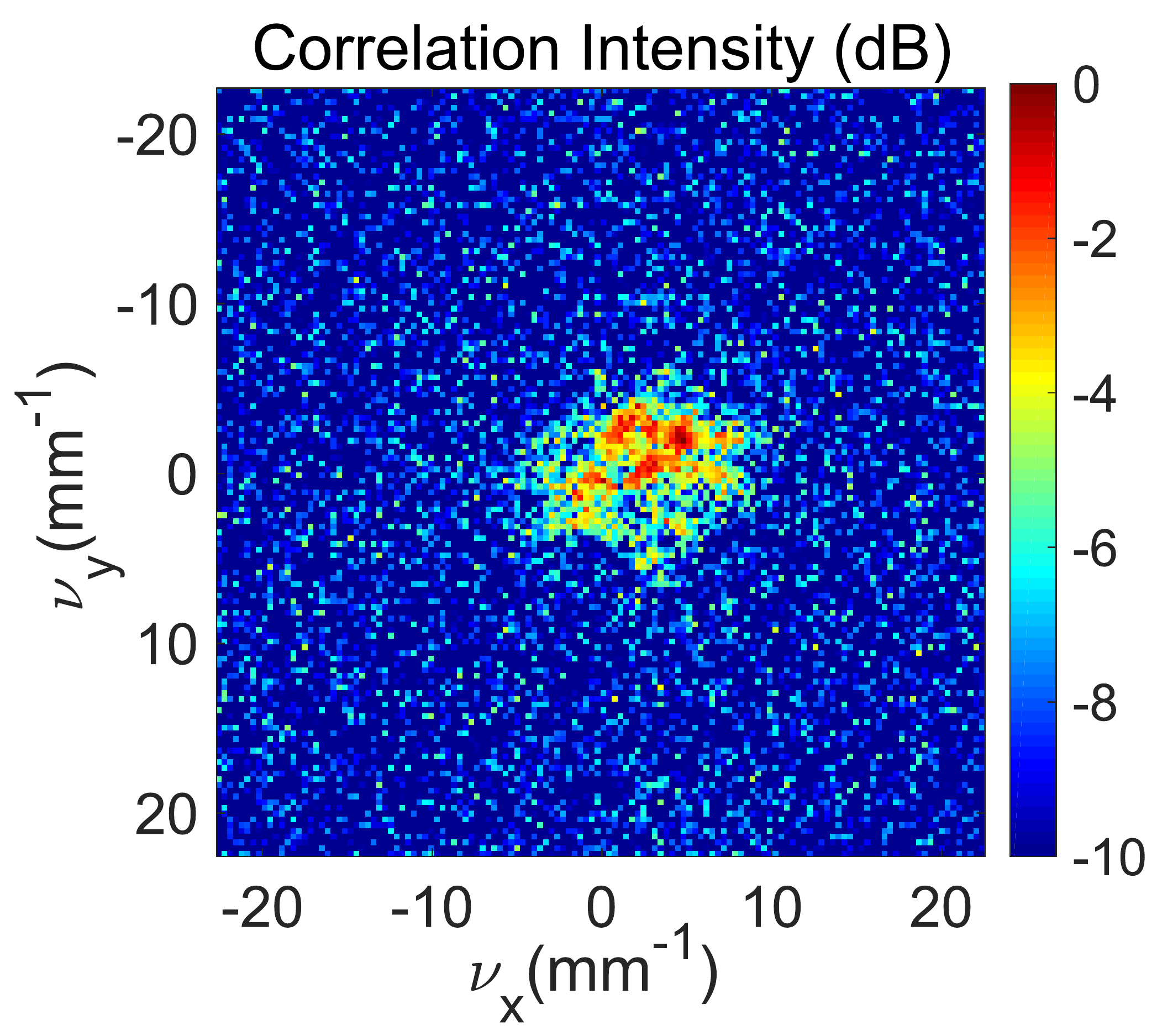}
	\caption{Speckle-like FF correlation function obtained when $D_1$ is removed in the setup depicted in Fig. \ref{fig3}}\label{twoD1diff}
\end{figure}
   
We first present one-dimensional computations in the FF of a type-1 crystal at degeneracy, by assuming that the birefringence compensates exactly the dispersion, ensuring perfect collinear phase-matching. Fig. \ref{unD}a presents the well known quasi-perfect anti-correlation of the signal and idler photons, corresponding to opposite wave-vectors. We have verified other well known features, like a shortening of the phase-matching range for a higher crystal length, or widening of the line for a narrower pump beam.  Fig. \ref{unD}b shows that inserting a pure phase scatterer in the FF has no effect, since it is equivalent to changing the phase of the biphoton wave function, with no consequence on the joint probability. This is the experimental situation reported in Fig. \ref{fig4}d, with no enlarging of the correlation peak. On the other hand, Fig. \ref{unD}c shows that, when inserting a scatterer in the NF, the phase matching line becomes a speckle with lines parallel to the  phase-matching line, as shown experimentally in Fig. 4a of Ref \cite{peeters_observation_2010} by recording temporal coincidences with punctual detectors. With scatterers in both planes, Fig. \ref{unD}d, there are no more lines, in good agreement with the Fig. 4c of this reference.
  
Second, we present, still in the FF, 2-D results for a type-2 BBO crystal of the same 0.8 mm length as in the experiment. Note however that, because of the quite small $32\times 32$ number of pixels, the lateral dimensions are smaller than in the experiment, with a FWHM width of the pump beam equal to 24 $\mu$m.  In  Fig. \ref{twoD}a, which shows the sum of the signal and idler intensities, we see that non critical type-2 phase-matching occurs around signal-idler directions that are shifted each-other from the walk-off angle \cite{lantz_phase-mismatch_2000} Without scatterer, we retrieve in Fig. \ref{twoD}b the narrow peak of the correlation function between the signal image and the 180\ensuremath{^\circ} rotated idler image, Fig. \ref{fig4}b in the experiment. Fig. \ref{twoD}c shows the speckle-like pattern obtained, with a thin scatterer in the NF, on the correlation between the signal and the rotated idler. The experimental speckle-like pattern (Fig. \ref{twoD1diff}) obtained first in \cite{gnatiessoro_imaging_2019} is similar, and both results can be explained by the transformation of the phase-matching pattern in speckle shown for the 1-D case in Fig. \ref{unD}c. If the correlation is realized without rotation, we obtain a beam without speckle structure, Fig. \ref{twoD}d, meaning that the photons do arrive by pairs, but without analogy with a coherent speckle. Last,  Fig. \ref{twoD}e and. Fig. \ref{twoD}f are obtained for scatterers in both the NF and the FF planes. Because of the addition of speckles generated for $r_s+r_i$ as well as for $r_s-r_i$ (see Fig. \ref{unD}d), we see only a whole beam structure, with some deformations.

The above-described method has clear advantages and drawbacks. For a pure biphoton state, it gives the full wave function, i.e. the most complete information, from which all statistical properties, like means and (co)variances, can be easily retrieved. Another advantage comes from the volume of calculations: a  simple integral for each value leads to a volume proportional to $N^3$, if N is the total of number of pixels in an unidimensional or bidimensional image. The main drawback comes from the assumption of independence between pairs. With this formalism, it is not possible to describe phenomena like squeezing, bunching or antibunching, that occur in optical  parametric amplification for higher gains, where the generation of a pair can seed the generation of another. To describe these phenomena, the two quadratures of the field must be taken into account. At high gains, stochastic simulations based on the Wigner formalism  \cite{lantz_spatial_2004,t-mills_simulating_2020} are certainly the most efficient, with a calculation volume proportional to $R \times N $, where $R$ is the number of repetitions of the simulation. At high gain, the corrections that allow the retrieval of the normally-ordered operators from the symmetrized ones can be neglected. In this case, it appears that one occurence of the simulation reproduces the main qualitative features, including fluctuations, of a single experimental shot \cite{lantz_numerical_2001} Nevertheless, from a theoretical point of view, only averages of a great number of simulations make sense, and this number can be huge at very small gain, where the effects of the introduced input quantum noise must be strongly averaged \cite{blanchet_purely_2010}. Fortunately, this regime corresponds to the conditions where the method developed in the present paper is accurate.\\ The Green function method, developed in \cite{treps_transverse_2000,lantz_spatial_2004} gives good results whatever the gain, for a computation time proportional to $N^3$, like the present method. Both the stochastic and the Green function methods do not give access to the entire wave-function, that has anyway a prohibitive number of elements for an image with several photons per pixel, but allow the calculation of the statistical features of the images. To summarize, the three methods have complementary validity domains. The calculation of the biphoton wave function, developed here, gives a complete information, i.e. probability amplitudes, for very small gain, with  a computation time which scales as $N^3$, $N$ being the total number of pixels. At high gain, a stochastic simulation is much more rapid, with a computation time proportional to $N$, and gives access to the statistical features as well as to the appearance of a single experiment. The Green function method is valid whatever the gain, with a computation time proportional to $N^3$ and results that give the statistical features of the images.

We have presented here first an extension to thick diffusing media of our previous work of correlation imaging of entangled photons traversing a diffusing medium. Second, we have shown that a simple numerical method allows the biphoton wave-function to be retrieved, with results in good agreement with the experiment. The immediate extension, in progress, would be a full characterization of the pixel to pixel coincidence speckles, by using no more a global correlation function. We expect, in agreement with the theory of Ref. \cite{beenakker_two-photon_2009} to retrieve the good contrast that disappears  when averaging over all speckles with the same coordinates difference or sum. Another perspective would be using more realistic thick media to characterize at which extent entanglement survives, as theoretically studied in \cite{cande_transmission_2014}. Last, control of scattering by using a spatial light modulator \cite {defienne_adaptive_2018} opens important perspectives.

\section*{Aknowledgments}
This work was partly supported by the French "Investissements d'Avenir" program, project ISITE-BFC (contract ANR-15-IDEX-03) and the RENATECH network and its FEMTO-ST MIMENTO technological facility.
\section*{Data availability}
The data that support the findings of this study are available from the corresponding author upon reasonable request.

\nocite{*}
\bibliography{biblioSoro}% Produces the bibliography via BibTeX.

\end{document}